\documentclass{fundam}

\usepackage[utf8]{inputenc}
\usepackage[T1]{fontenc}
\usepackage{acronym}
\usepackage{dsfont}
\usepackage{color}
\usepackage{subfigure}
\usepackage{epsfig}
\usepackage{graphics}
\usepackage{graphicx}
\usepackage{epstopdf}
\usepackage{url}
\usepackage{times}
\usepackage[english,ruled]{algorithm2e}
\usepackage{newtxtext,newtxmath}

\usepackage{listings}
\lstdefinelanguage{its}
{morekeywords={if, =, +, >=, +, -, (, ), [, ], true, false, typedef, transition, int,GAL, abort, !, \{, \}, label, ", &&, ., composite, synchronization,for,gal, array},
morecomment=[l]{//},
sensitive=false,
}
\newcommand{\nat}%
{\ensuremath{\mathds{N}}}
\newcommand{\zrel}%
{\ensuremath{\mathds{Z}}}
\newcommand{\rea}%
{\ensuremath{\mathds{R}}}
\newcommand{\bool}%
{\ensuremath{\mathds{B}}}

\newcommand{\rational}%
{\ensuremath{\mathds{Q}}}

\newcommand{\tuple}[1]{\langle #1 \rangle}

\newcommand{\Places}{\ensuremath{\mathcal{P}}}
\newcommand{\Trans}{\ensuremath{\mathcal{T}}}
\newcommand{\Pre}{\ensuremath{\mathcal{W}_-}}
\newcommand{\Post}{\ensuremath{\mathcal{W}_+}}
\newcommand{\Effect}{\ensuremath{\mathcal{W}_e}}
\newcommand{\tPre}{\ensuremath{\mathcal{W}_-^\top}}
\newcommand{\tPost}{\ensuremath{\mathcal{W}_+^\top}}
\newcommand{\Preset}[1]{\ensuremath{\bullet{#1}}}
\newcommand{\Postset}[1]{\ensuremath{{#1}\bullet}}

\newcommand{\Reach}{\ensuremath{\mathcal{R}}}
\newcommand{\Stutter}{\ensuremath{\mathcal{S}_{t}}}

\newcounter{rrule}
\newenvironment{rrule}[1][]{\refstepcounter{rrule}\par\medskip
   \textbf{Rule~\therrule. #1} \rmfamily}{\medskip}

\renewcommand{\implies}{\Rightarrow}
\renewcommand{\impliedby}{\Leftarrow}

  \usepackage{hyperref}

\begin{document}

%
\setcounter{page}{319}
\publyear{2021}
\papernumber{2090}
\volume{183}
\issue{3-4}

  \finalVersionForARXIV


\title{Symbolic and Structural Model-Checking}

\author{Yann Thierry-Mieg\thanks{Address  for correspondence: Sorbonne Université, CNRS, LIP6,  F-75005 Paris, France}
\\
Sorbonne Université, CNRS, LIP6,\\
F-75005 Paris, France \\
Yann.Thierry-Mieg@lip6.fr
}


\maketitle              

\runninghead{Y. Thierry-Mieg}{Symbolic and Structural Model-Checking}

\vspace*{-4mm}
\begin{abstract}
Brute-force model-checking consists in exhaustive exploration
of the state-space of a Petri net, and meets the dreaded state-space explosion problem.

In contrast, this paper shows how to solve model-checking problems using a combination of techniques
that stay in complexity proportional to the size of the net structure rather than to the state-space size.

We combine an SMT based over-approximation to prove that some behaviors are unfeasible, an under-approximation using memory-less sampling of runs to find witness traces or counter-examples, and a set of structural reduction rules that can simplify both the system and the property.

This approach was able to win by a clear margin the model-checking contest 2020 for reachability queries as well as deadlock detection, thus demonstrating the practical effectiveness and general applicability of the system of rules presented in this paper.

\end{abstract}

\section{Introduction}

Given a Petri net and an assertion on markings of places of the net, we consider the problem of proving that
the assertion is an \textit{invariant} that holds in all reachable markings. We introduce a combination of three solution strategies: $1.$~we try to prove the invariant holds using a system of SMT constraints to symbolically \emph{over-approximate} reachable states $2.$~we try to disprove the invariant by using a memory-less exploration that can randomly or with guidance encounter counter-example states thus \emph{under-approximate} the behavior,  and $3.$~we use structural reduction rules that preserve the properties of interest through a transformation that reduces the model size. These approaches reinforce and complement each other to provide a powerful symbolic and structural model-checking procedure.
It has a relatively low complexity since it is related to the size of the net rather than to the size of the state~space.

We consider an over-approximation of the state space, symbolically represented as set of constraints over a system of variables encoded in an SMT solver~\cite{z308}. SMT solvers are a modern technology that offer
the benefits of both Integer Linear Programming (ILP) solvers (that have been used for a similar purpose in e.g.~\cite{Wolf11}) and SAT solvers for more flexibility in the expression of constraints.
We thus use this SMT based over-approximation of reachable markings to detect unfeasible behavior.

The SMT procedure while powerful is only a semi-decision procedure in the sense that UNSAT answers \emph{prove that the invariant} holds ($\lnot \mathcal{I}$ is \emph{not} reachable), but SAT answers are not trusted because we work with an over-approximation of the system.
A SAT answer does however suggest a candidate solution that contradicts the invariant. In the spirit of a CEGAR scheme~\cite{cegar00} or of the approach of~\cite{Wolf11}, we can try to replay this candidate on the real system to exhibit a concrete counter-example thus proving the invariant \emph{does not hold}.


We thus develop a sampling based approach that under-approximates the state-space. This consists in a memory-less and fast transition engine able to explore up to millions of states per second. This engine can run in pseudo-random exploration mode or can be guided by a candidate solution coming from the SMT engine. If it can find a reachable marking that does not satisfy the target assertion, the invariant is disproved.

We combine these solutions with a set of structural reduction rules, that can simplify the net by examining its structure, and provide a smaller net where parts of the behavior are removed or accelerated over while preserving properties of interest.
Structural reductions can be traced back at least to Lipton's transaction reduction~\cite{Lipton75} and in the context of Petri nets to Berthelot's seminal paper~\cite{Berthelot85}.
 Structural reductions are complementary of any other verification or model-checking strategies, since they build a simpler net and property that can be further analyzed using other methods.
 Reducing the size of the net considerably helps both the SMT based solutions because there are less variables and constraints and the pseudo-random sampling as it is more likely to find counter-examples if the state space is small.

The paper~\footnote{This paper is an extended and revised version of~\cite{SRRPN20}, with new relaxed conditions for "partial" agglomeration in section 4.3, an updated expression of SMT constraints to avoid using quantifiers, and a new experimental section 5.2.} is organized as follows, section~$2$ presents the notations we use, section~$3$ presents the over-approximation using SMT. We then present a set of $22$ structural reduction rules in section $4$. Section~$5$ presents detailed experimental analysis of the excellent results obtained by the tool implementing this approach in the model-checking contest~\cite{mcc2019} in 2020.



\section{Definitions}
\label{secDefs}

\subsection{Petri net syntax and semantics}

\begin{definition}
Structure. A Petri net $N=\tuple{\Places,\Trans,\Pre,\Post, m_0}$ is a tuple where \Places{} is the set of places, \Trans{} is the set of transitions, $\Pre: \Places \times \Trans \mapsto \nat$ and $\Post: \Places \times \Trans \mapsto \nat$ represent the pre and post incidence matrices, and $m_0: \Places \mapsto \nat$ is the initial marking.
\end{definition}

Notations:
We use $p$ (resp. $t$) to designate a place (resp. transition) or its index dependent on the context. We let markings $m$ be manipulated as vectors of natural with $|\Places|$ entries. We let $\Pre(t)$ and $\Post(t)$ for any given transition $t$ represent vectors with $|\Places|$ entries. $\tPre, \tPost$ are the transposed flow matrices, where an entry $\tPre(p)$ is a vector of $|\Trans|$ entries. We note $\Effect = \Post - \Pre$ the integer matrix representing transition \textit{effects}.

In vector spaces, we use $v \geq v'$ to denote $\forall i, v(i)\geq v'(i)$, and we use sum $v+v'$ and scalar product $k \cdot v$ for scalar $k$ with usual element-wise definitions.

We note $\Preset{n}$ (resp. \Postset{n}) the pre set (resp. post set) of a node $n$ (place or transition). E.g. for a transition $t$ its pre set is $\Preset{t}= \{ p \in \Places \mid \Pre(p,t) > 0 \}$.  A marking $m$ is said to \textit{enable} a transition $t$ if and only if $m \geq \Pre(t)$. A transition $t$ is said to \emph{read} from place $p$ if $\Pre(p,t)>0 \land \Effect(p,t)=0$.

\vspace{-0.5em}
\begin{definition}
Semantics. The semantics of a Petri net are given by the firing rule $\xrightarrow{t}$ that relates pairs of markings: in any marking $m \in  \nat^{|\Places|}$, if $t \in \Trans$ satisfies $m \geq \Pre(t)$, then $m\xrightarrow{t}m'$ with $m' = m + \Post(t) - \Pre(t)$. The reachable set $\Reach{}$ is inductively defined as the smallest subset of $\nat^{|\Places|}$ satisfying $m_0 \in \Reach$, and $\forall t \in \Trans{}, \forall m \in \Reach{}, m \xrightarrow{t} m' \implies m' \in \Reach$.
\end{definition}

\vspace{-1em}
\subsection{Properties of interest}

We focus on deadlock detection and verification of safety properties. A net contains a deadlock if its reachable set contains a marking $m$ in which no transition is enabled.
A safety property asserts an invariant $\mathcal{I}$ that all reachable states must satisfy. The invariant is given as a Boolean combination ($\lor, \land, \lnot$) of atomic propositions that can compare ($\bowtie \in \{<, \leq, =, \geq, >\}$) arbitrary weighted sum of place markings to another sum or a constant, e.g. $\sum_{p \in \Places} \alpha_p \cdot m(p) \bowtie k$, with $\alpha_p \in \zrel$ and $k \in \zrel$.

\medskip
In the case of safety properties, the \textit{support} of a property is the set of places whose marking is truly used in the predicate, i.e. such that at least one atomic proposition has a non zero $\alpha_p$ in a sum. The support $\mathit{Supp} \subseteq \Places{}$ of the property defines the subset $\Stutter{} \subseteq \Trans$ of \textit{invisible} or \textit{stuttering} transitions $t$ satisfying $\forall p \in \mathit{Supp}, \Effect(p,t)=0$ \footnote{This sufficient condition for being stuttering might be relaxed by a more refined examination of the property and the effect of transitions on its truth value with respect to its value initially.}. For safety, we are only interested in the projection of reachable markings over the variables in the support, values of places in $\Places{}\setminus \mathit{Supp}$ are not \emph{observable} in markings. A small support means more potential reductions, as rules  mostly cannot apply to observed places or their neighborhood.

\section{Proving with SMT constraints}
\label{secSMT}

In this section, we define an over-approximation of the state space, symbolically manipulated as a set of constraints over a system of variables encoded in a Satisfiability Modulo Theory (SMT) solver~\cite{z308}. We use this approximation to detect unfeasible behavior. We present constraints that can be progressively added to a solver to over-approximate the state space with increasing accuracy. Structural reduction rules based on a behavioral characterization of target conditions are also possible in this context (see Sec~.\ref{SMTRed}).

\subsection{Approximating with SMT}

SMT solvers are a powerful and relatively recent technology that enables flexible modeling of constraint satisfaction problems using first order logic and rich data types and theories, as well as their combinations. We use both linear arithmetic over reals and integers (LRA and LIA) to approximate the reachable set of states by constraints over variables representing the marking of places.

As first step in all approaches, we define for each place $p \in \Places{}$ a variable $m_p$ that represents its marking. These variables are initially only constrained to be positive: $\forall p \in \Places{}, m_p \geq 0$. If we know that the net is one-safe (all place markings are at most one) e.g. because the net was generated, we add that information: $\forall p \in \Places{}, m_p \leq 1$.

We then suppose that we are trying to find a reachable marking that \emph{invalidates} a given invariant property $I$ over a support. In other words we assert that the $m_p$ variables satisfy $\lnot I$. For deadlocks, we consider
the invariant $I$ asserting that at least one transition is enabled, expressed in negative form as $\lnot I = \forall t \in \Trans{}, \exists p \in \Preset{t}, m_p < \Pre(p,t)$, and thus reduce the Deadlock problem to Safety.

An UNSAT answer is a definitive "NO" that ensures that $I$ is indeed an invariant of the system. A SAT answer provides a candidate marking $m_c$ with an assignment for the $m_p$ variables, but is unreliable since we are considering an over-approximation of the state-space.
To define the state equation constraint (see below) we also add a variable $n_t$ for each transition $t$ that counts the number of times it fired to reach the solution state. This Parikh count provides a guide for the random explorer.

\medskip
\textbf{Workflow:}
Because we are hoping for a definitive UNSAT answer,
and that we have a large set of constraints to approximate the state space, they are fed incrementally to the solver, hoping for an early UNSAT result. In practice we check satisfiability after every few assertions, and start with the simpler constraints that do not need additional variables.

Real arithmetic is much easier to solve than integer arithmetic, and reals are an over-approximation of integers since if no solution exists in reals (UNSAT), none exists in integers either.
We therefore always first incrementally add constraints using the real domain, then at the end of the procedure, if the result is still SAT, we test if the model computed (values for place markings and Parikh count) are actually integers. If not, we escalate the computation into integer domain, restarting the solver and similarly adding the simplest constraints first.
At the end of the procedure we either get UNSAT or a "likely" Parikh vector that can be used to cheaply test if the feasibility detected by SAT answer is actually doable.

\subsection{Incremental constraints}

We now present the constraints we use to approximate the state space of the net, in order of complexity and induced variables that corresponds to the order they are fed to our solver. We progressively add generalized flow constraints, trap constraints, the state equation, read arc constraints, and finally add new "causality" constraints.

\subsubsection{Generalized flows}

A flow $\mathcal{F}$ is a weighted sum of places $\sum_{p\in \Places{}} \alpha_p \cdot m_p$ that is an invariant of the state space, it has a constant value in all markings that can be computed as $\sum_{p\in \Places{}} \alpha_p \cdot m_0(p)$ using the initial marking. A semi-flow only has positive integers as $\alpha_i \in \nat$ coefficients while generalized flows may have both positive and negative integer coefficients $\alpha_i \in \zrel$.

It is possible to compute a generator basis of all generalized flows of a net in polynomial time and space, and the number of flows obtained is itself polynomial in the size of the net. We use for this purpose a variant of the algorithm described in~\cite{flow88}, initially adopted from the code base of~\cite{APT15} and then optimized for our scenario. This provides a polynomial number of simple constraints each only having as variables a subset of places (the $\alpha_i$ are fixed). The constant value of the flow in any marking can be deduced from the initial marking of the net.
We do not attempt to compute a basis of semi-flows as there can be an exponential number of them, but we still first assert semi-flow constraints (if any were found) before asserting generalized flows. A semi-flow has far fewer solutions due to markings being positive, so these are "easy" constraints for the solver to process. For instance $m(p)+m(p')=K$ has $K+1$ solutions, and puts a structural bound on the marking of $p$ and $p'$, while $m(p)-m(p')=K$ admits an infinite set of solutions.\vspace*{-1mm}

\subsubsection{Trap constraints}

In \cite{Esparza00}, to reinforce a system of constraints on reachable states, "trap constraints" are proposed. Such a constraint asserts that an initially marked trap must remain marked.

\begin{definition}
Trap. A trap $S$ is a subset of places such that any transition consuming from the set must also feed the set.  $S \subseteq \Places{}$ is a
\emph{trap} iff.\vspace*{-1mm}
$$
\forall p \in S, \forall t \in \Postset{p}, \exists p' \in \Postset{t} \land p' \in S.\vspace*{-1mm}
$$
\end{definition}

The authors of \cite{Esparza00} show that traps provide constraints that are a useful
complement to state equation based approaches, as they can discard unfeasible behavior that is otherwise compatible with the state equation. The problem is that in general these constraints are worst case exponential in number. Leveraging the incremental nature of SMT solvers, we therefore propose to only introduce "useful" trap constraints, that actually contradict the current candidate solution.

We consider the candidate state $m_c$ produced as SAT answer to previous queries, and try to contradict it using a trap constraint: we look for an initially marked trap that is empty in the solution. The search for such a potential trap can be done using a separate SMT solver instance.

For each place $p$ in the candidate state $m_c$, we introduce a Boolean variable $b_p$ that will be true when $p$ is in the trap. We then add the trap constraints:\medskip

\begin{tabular}{@{}ll@{}}
$\bigvee_{(p \in \Places \mid m_0(p) > 0)} b_p$  & trap was initially marked \\
$\forall p \in \Places, b_p \implies \bigwedge_{t \in \Postset{p}} ( \bigvee_{p' \in \Postset{t}} b_{p'}$) & trap definition \\
$\bigwedge_{(p \in \Places \mid m_c(p) > 0)} \lnot b_p$ & only consider unmarked places in our candidate\medskip
\end{tabular}

If this problem is SAT, we have found a trap $S$ from which we can derive a  constraint expressed as $\bigvee_{p \in S} m_p > 0$ that can be added to the main solution procedure. Otherwise, no trap constraint existed that could contradict the given witness state. The procedure is iterated until no more useful trap constraints are found or UNSAT is obtained.

\vspace{-1em}
\subsubsection{State equation}

The state equation~\cite{stateEq77} is one of the best known analytical approximations of the state space of a Petri net.

\begin{definition}
State equation. We define for each transition $t \in \Trans{}$ a variable $n_t \in \nat$. We assert for each place $p \in \Places$ that
$$
m_p = m_0(p) + \sum_{t \in \Postset{p} \cup \Preset{p}} n_t\cdot \Effect(p,t)
$$
\end{definition}

The state equation constraint is thus implemented by adding for each transition $t\in \Trans{}$ a variable $n_t \geq 0$ and then asserting for each place $p \in \Places{}$ a linear constraint. Instead of considering all transitions and adding a variable for each of them, we can limit ourselves to one variable per possible transition \emph{effect}, thus having a single variable for transitions $t,t'$ such that $\Effect(t)=\Effect(t')$. Care must be taken when interpreting the resulting Parikh vectors however.
In the worst case, the state equation adds $|\Trans{}|$ variables and $|\Places{}|$ constraints, which can be expensive for large nets. We always start by introducing the constraints bearing on places in the support.
Although flow constraints are redundant with respect to the state equation (the state equation implies the flow constraints), leveraging the incremental nature of the solver we do not restart it or clear the flow constraints however.

A side effect of introducing the state equation constraints is that we now have a candidate Parikh firing vector, in the form of the values taken by $n_t$ variables. The variables in this Parikh vector can now be further constrained, as we now show.

\vspace{-0.6em}
\subsubsection{Read $\implies$ feed constraints}
A known limit of the state equation is the fact that it does not approximate
read arc behavior very well, since it only reasons with actual effects $\Effect$ of transitions.
However we can further constrain our current solution to the state equation by requiring that for any transition $t$ used in the candidate Parikh vector, that reads from an initially insufficiently marked place $p$, there must be a transition $t'$ with a positive Parikh count that feeds $p$.

\begin{definition}
Read Arc Constraint. For each transition $t \in \Trans{}$, for every initially insufficiently marked place it reads from i.e. $\forall p \in \Preset{t}$, such that $\Effect(p,t)=0 \land \Pre(p,t) > m_0(p)$, we assert that:
$$
n_t > 0 \implies \bigvee_{ \{t' \in \Preset{p} \setminus \{t\}\ \mid  \Effect(p,t') > 0\}}  n_{t'}>0
$$
\end{definition}

These read arc constraints are easy to compute and do not introduce any additional variables so they can usually safely be added after the problem with the state equation returned SAT, thus refining the solution.

While in practice it is rare that these additional constraints allow to conclude UNSAT, they frequently improve the feasibility of the Parikh count solution on nets that feature a lot of read arcs (possibly due to reductions), going from an unfeasible solution to one that is in fact realizable.
The more general causality constraints below subsume the effect of these constraints, but these constraints are simpler and do not introduce additional variables, so we introduce them first.

\subsubsection{Causality constraints}
Solutions to the state equation may contain transition cycles, that "borrow" non-existing tokens and then return them. This leads to spurious solutions
that are not in fact feasible. However, we can break such cycles of transitions if we consider the partial order that exists over the first occurrence of each transition in a potential concrete trace realizing a Parikh vector.

Indeed, any time a transition $t$ consumes a token in place $p$, but $p$ is not initially sufficiently marked, it must be the case that there is another transition $t'$ that feeds $p$, and that $t'$ \emph{precedes} $t$ in the trace.

We thus consider a precedes $\prec \subseteq \Trans{} \times \Trans{}$ relation between transitions that is : non-reflexive $\forall t \in \Trans{}, \lnot (t \prec t$), transitive $\forall t_1,t_2,t_3 \in \Trans{}, t_1 \prec t_2 \land t_2 \prec t_3 \implies t_1 \prec t_3$,
anti-symmetric $\forall t_1,t_2 \in \Trans, t_1 \prec t_2 \implies \lnot (t_2 \prec t_1$). This relation defines a strict partial order over transitions.

\begin{definition}
Causality Constraint.
We add the definition of the precedes relation to the solver.
For each transition $t \in \Trans{}$, for each input of its input places that is insufficiently marked i.e. $\forall p \in \Preset{t},
\Pre(p,t) > m_0(p)$, we assert that
$$
n_t > 0 \implies \bigvee_{ \{t' \in \Preset{p} \setminus \{t\}\ \mid  \Effect(p,t') > 0\}}  ( n_{t'}>0 \land t' \prec t )
$$
\end{definition}

These constraints reflect the fact that insufficiently marked places must be fed \emph{before} a continuation can take place. These constraints offer a good complement to the state equation as they forbid certain Parikh solutions that use a cycle of transitions and "borrow" tokens:
such an Ouroboros-like cycle now needs a causal predecessor to be feasible.
Our solutions still over-approximate
the state-space as we are only reasoning on the first firing of each transition, and we construct conditions for each predecessor place separately, so we cannot guarantee that all input places of a transition have been \emph{simultaneously} marked.\\
\textbf{Notes:}The addition of causal constraints forming a partial order to refine state equation based reasoning has not been proposed before in the literature to our knowledge.

To encode the \emph{precedes} constraints in an SMT solver the approach we found most effective in practice consists in defining a new integer (or real) variable $o_t$ for each transition $t$, and use strict inferior $o_{t_1} < o_{t_2}$ to model the precedes relation $t_1 \prec t_2$. This remains a partial order as some $o_t$ variables may take the same value, and avoids introducing any additional theories or quantifiers.


\vspace{-.5em}
\section{Structural reduction rules}
\label{structuralRed}
We are given a Petri net $N$ and either a set of safety invariants or a deadlock detection query. We consider the problem of building a structurally smaller net $N'$ and/or simpler properties such that the resulting properties hold on the smaller net if and only if the original properties hold on the original net.

This section defines a set of $22$ structural reduction rules.
The main idea in reduction rules is either to discard parts of the net or to accelerate over parts of the behaviors by fusing adjacent transitions. Reduction rules exploit the locality property of transitions to define a reduction's effect in a small neighborhood.
Most rules can be adapted to support preservation of stutter-invariant temporal logic.

Structural reductions have been widely studied with generalizations that apply to many other models than Petri nets e.g.~\cite{Laarman18}.
The classical reduction rules~\cite{Berthelot85} include pre and post agglomeration, for which~\cite{EHPP05,HPP06} give broad general definitions that can be applied also to colored nets.
More recently, several competitors in the Model Checking Contest have worked on the subject,~\cite{Srba19} defines $8$ reduction rules used in the tool Tapaal and~\cite{berthomieu19} defines very general transition-centric reduction rules used in the tool Tina.
Thus, while some of the more basic rules presented exist in the literature, most of the rules in this paper are presented with relaxed conditions that widen their application scope, or are a fully novel approach introduced in~\cite{SRRPN20}.


Deadlock detection could be stated as the invariant "at least one transition is enabled". But this typically implies that all places are in the support, severely limiting rule application. So instead we define deadlock specific reductions that consider that the support is empty, and that are mainly concerned with preserving divergent behavior (loops).

This section is split into five subsections, $4.1$ and $4.2$ introduce very basic and general rules for transitions and places, $4.3$ presents our agglomeration based-rules that widen the application scope of existing definitions of agglomeration, $4.4$ presents new rules that reason on a graph that is an abstraction of the structure of the net and can remove large and complex parts of a net, $4.5$ presents rules that leverage the SMT based over-approximation of reachable states to test behavioral application conditions.

For each rule, we give a name and identifier; whether it is applicable to deadlock detection, safety or both; an informal description of the rule; a formal definition of the rule; a sketch of correctness where $\implies$ proves that states observably satisfying the property (or deadlocks) are not removed by the reduction, $\impliedby$ proves that new observable states (or deadlocks) are not added.

\subsection{Elementary transition rules}\vspace*{-1mm}

\rrule{Equal transitions modulo $k$} \\
\textbf{Applicability:} Safety, Deadlock \\
\textbf{Description:} When two transitions  are equal modulo $k$, the larger one can be discarded. \\
\textbf{Definition:}
If
$
\exists t,t' \in \Trans, \exists k \in \nat, \Pre(t)=k \cdot \Pre(t') \land \Post(t)=k \cdot \Post(t')
$, discard $t$. \\
\textbf{Correctness:} $\implies:$ In any state where $t$ is enabled, $t'$ must be enabled, and firing $k$ times $t'$ leads to the same state as firing $t$. $\impliedby:$ Discarding transitions cannot add states. For deadlocks, any state enabling $t$ in the original net still has a successor by $t'$.
\smallskip

\rrule{Dominated transition} \\
\textbf{Applicability:} Safety, Deadlock \\
\textbf{Description:} When a transition $t$ has the same effect as $t'$ but more preconditions, which can happen due to read arc behavior, $t$ can be discarded. \\
\textbf{Definition:}
If $\exists t,t' \in \Trans, \Effect(t)=\Effect(t') \land \Pre(t) \geq \Pre(t'),$ discard $t$. \\
\textbf{Correctness:} $\implies:$ any state that enables $t$ also enables $t'$, and the resulting state is the same. $\impliedby:$ Discarding transitions cannot add states. For deadlocks, any state enabling $t$ in the original net still has successors by $t'$ in the resulting net.
\smallskip

\rrule{Redundant Composition} \\
\textbf{Applicability:} Safety, Deadlock \\
\textbf{Description:} When a transition $t$ has the same effect and more input places than a composition $t_1.t_2$, where $t_1$ enabled implies $t_1.t_2$ is enabled, $t$ can be discarded. \\
\textbf{Definition:}
If $\exists t,t_1,t_2 \in \Trans, \Pre(t) \geq \Pre(t_1), \Post(t_1) \geq \Pre(t_2), \Effect(t) = \Effect(t_1) + \Effect(t_2)$, discard $t$. \\
\textbf{Correctness:} $\implies:$ any state that enables $t$ also enables $t_1$, $t_2$ is necessarily enabled after firing $t_1$, and the state reached by the sequence $t_1.t_2$ is the same as reached by $t$. $\impliedby:$ Discarding transitions cannot add states. For deadlocks, any state enabling $t$ in the original net still has successors by $t_1$ in the resulting net. \\
\textbf{Notes:}
This pattern could be extended to compositions of more transitions, but there is a risk of explosion as we end up exploring intermediate states of the trace.~\cite{berthomieu19} notably proposes a more general version that subsumes the four first rules presented here but is more costly to evaluate. 
\smallskip

\rrule{Neutral transition} \\
\textbf{Applicability:} Safety \\
\textbf{Description:} When a transition has no effect it can be discarded. \\
\textbf{Definition:}
If $\exists t \in \Trans, \Pre(t)=\Post(t),$ discard $t$. \\
\textbf{Correctness:} $\implies:$ any state reachable by a firing sequence using $t$ can also be reached without firing $t$. $\impliedby:$ Discarding transitions cannot add states.
\smallskip

\rrule{Sink Transition} \\
\textbf{Applicability:} Safety \\
\textbf{Description:} A transition $t$ that has no outputs and is stuttering can be discarded. \\
\textbf{Definition:}
If $\exists t \in \Stutter, \Postset{t} = \emptyset$, discard $t$. \\
\textbf{Correctness:} $\implies:$ any state reached by firing $t$ enables less subsequent behaviors since tokens were consumed by $t$. Any firing sequence of the original net using $t$ is still possible in the new net if occurrences of $t$ are removed, and leads to a state satisfying the same properties because $t$ stutters.
$\impliedby:$ Discarding transitions cannot add states.
\smallskip

\rrule{Source Transition} \\
\textbf{Applicability:} Deadlock \\
\textbf{Description:} If a transition has no input places then the net has no reachable deadlocks. \\
\textbf{Definition:}
If $\exists t \in \Trans, \Preset{t} = \emptyset$, discard all places and discard all transitions  except $t$. \\
\textbf{Correctness:} $\implies:$ Since $t$ is fireable in any reachable state, there cannot be reachable deadlock states.
$\impliedby:$ The resulting model has no deadlocks, like the original model.

\vspace*{-1mm}
\subsection{Elementary place rules}

\rrule{Equal places modulo $k$} \\
\textbf{Applicability:} Deadlock, Safety \\
\textbf{Description:} When two places are equal modulo $k$ in terms of flow matrices
the one with the least tokens can be discarded. If both places are outside the support either one can be discarded otherwise the place not in the support can still be discarded.  \\
\textbf{Definition:}
If $\exists p \in \Places\setminus \mathit{Supp}, \exists p' \in \Places, \exists n \in \nat, \exists k \in \rational, k=n \lor k=\frac{1}{n}, m_0(p) \geq k \cdot m_0(p'), \tPre(p)=k \cdot \tPre(p') \land \tPost(p)=k \cdot \tPost(p'),$ discard $p$. \\
\textbf{Correctness:}  $\implies:$ Removing a place cannot remove any behavior. $\impliedby:$
Inductively we can show that $m(p) \geq k\cdot m(p')$ in any reachable marking $m$. Thus enabling conditions on output transitions $t$ of $\Postset{p}=\Postset{p'}$ are always limited by the marking of $p'$: either $p'$ is insufficiently marked or both $p$ and $p'$ are sufficiently marked to let $t$ fire. Removing the condition on $p$ thus does not add any behavior.
\smallskip

\rrule{Sink Place} \\
\textbf{Applicability:} Deadlock, Safety \\
\textbf{Description:} When a place $p$ has no outputs, and is not in the support of the property, it can be removed. \\
\textbf{Definition:}
If $\exists p \in \Places\setminus \mathit{Supp}, \Postset{p}=\emptyset$, discard $p$. \\
\textbf{Correctness:} $\implies:$ Removing a place cannot remove any behavior. $\impliedby:$ Since the place had no outputs it already could not enable any transition in the original net.
\smallskip

\rrule{Constant place} \\
\textbf{Applicability:} Deadlock, Safety \\
\textbf{Description:} When a place's marking is constant (typically because of read arc behavior), the place can be removed and the net can be simplified by "evaluating" conditions on output transitions. \\
\textbf{Definition:}
If $\exists p \in \Places, \tPre(p)=\tPost(p)$, discard $p$ and any transition $t \in \Trans$ such that  $\Pre(p,t) > m_0(p)$. \\
\textbf{Correctness:}  $\implies:$ Removing a place cannot remove any behavior.
The transitions discarded could not be enabled in any reachable marking so no behavior was lost.
$\impliedby:$
The remaining transitions of $\Postset{p}$ have one less precondition, but it evaluated to true in all reachable markings, so no behavior was added. Discarding transitions cannot add states.\\
\textbf{Notes:} This reduction also applies to places in the support, leading to simplification of the related properties.
\smallskip

\rrule{Maximal Unmarked Siphon} \\
\textbf{Applicability:} Deadlock, Safety \\
\textbf{Description:} An unmarked \textit{siphon} is a subset of places that are not initially marked and never will be in any reachable state. These places can be removed and adjacent transitions can be simplified away. \\
\textbf{Definition:}
A maximal unmarked siphon $S \subseteq \Places$ can be computed by initializing with the set of initially unmarked places $S = \{p \in \Places \mid  m_0(p)=0 \}$, and $T \subseteq \Trans$ with the full set $\Trans$ then iterating:
\begin{itemize}
    \item Discard from $T$ any transition that has no outputs in $S$, $t \in T, \Postset{t} \cap S =\emptyset$,
    \item Discard from $T$ any transition $t$ that has no inputs in $S$ and discard all of $t$'s output places from $S$. So $\forall t \in T,$ if $\Preset{t} \cap S=\emptyset$, discard $t$ from $T$ and discard $\Postset{t}$ from $S$,
\item iterate until a fixed point is reached.
\end{itemize}
If $S$ is non-empty, discard any transition $t$ such that $\Preset{t} \cap S \neq \emptyset$ and all places in $S$. \\
\textbf{Correctness:} $\implies:$ The discarded transitions were never enabled so no behavior was lost. Removing places cannot remove behavior. $\impliedby:$ Removing transitions cannot add behavior. The places removed were always empty so they could not enable any transition.\\
\textbf{Notes:} Siphons have been heavily studied in the literature~\cite{siphon16}. This reduction also applies to places in the support, leading to simplification of the related properties.
\smallskip

\rrule{Bounded Marking Place} \\
\textbf{Applicability:} Deadlock, Safety \\
\textbf{Description:} When a place $p$ has no true inputs, i.e. all transitions effects can only reduce the marking of $p$, $m_0(p)$ is an upper bound on its marking that can be used to reduce adjacent transitions. \\
\textbf{Definition:}
If $\exists p \in \Places, \forall t \in \Trans, \Effect(p,t) \leq 0$, discard any transition $t' \in \Trans$ such that $\Pre(p,t') > m_0(p)$. \\
\textbf{Correctness:} $\implies:$ Since the transitions discarded were never enabled in any reachable marking, removing them cannot lose behaviors. $\impliedby:$ Discarding transitions cannot add behaviors.
\smallskip

\rrule{Implicit Fork/Join place} \\
\textbf{Applicability:} Deadlock, Safety \\
\textbf{Description:} Consider a place $p$ not in the support that only touches two transitions: $t_{fork}$ with two outputs (including $p$) and $t_{join}$ with two inputs (including $p$). If we can prove that the only tokens that can mark the other input $p'$ of $t_{join}$ must result from firings of $t_{fork}$, $p$ is implicit and can be discarded. \\
\textbf{Definition:}
If~$\exists p \in \Places\setminus \mathit{Supp}, \exists t_f,t_j \in \Trans, \Preset{p}=\{t_f\} \land \Postset{p}=\{t_j\}, \Post(p,t_f)=\Pre(p,t_j)=1, \Postset{t_f}=\{p,p''\} \land \Post(p'',t_f)=1, \Preset{t_j}=\{p,p'\} \land \Pre(p',t_j)=1,$ then if $p'$ is \textit{induced by} $t_f$, discard $p$. \\
We use a simple recursive version providing sufficient conditions for the test "is $p'$ induced by $t$":
\begin{itemize}
\itemsep=0.9pt
    \item If $p'$ has $t$ as single input and $\Post(p',t)=1$, return true.
    \item If $p'$ has a single input $t'$ and $\Post(p',t')=1$, if there exists any input $p''$ of $t'$, such that $\Post(p'',t')=1$, and (recursively) $p''$ is induced by $t$ return true. Else false.
\end{itemize}
\textbf{Correctness:}  $\implies:$ Removing a place cannot remove any behavior.
$\impliedby:$ In any marking that disabled $t_j$, either both $p$ and $p'$ were unmarked, or only $p'$ was unmarked. The constraint on weights being one all along the "induced by" path avoids token confusion, and ensures the number of tokens reaching $p'$ is consistent with number of tokens in $p$. Removing the condition on $p$ thus does not add behavior. \\
\textbf{Notes:}
There are many ways we could refine the "induced by" test, and widen the application scope, but the computation should remain fast. We opted here for a reasonable complexity vs. applicability trade-off. The implementation further bounds recursion depth (to $5$ in the experiments), and protects against recursion on a place already in the stack. Implicit places are studied in depth in~\cite{colom99}, the concept is used again in SMT backed Rule~$21$.
\smallskip

\rrule{Future equivalent place} \\
\textbf{Applicability:} Deadlock, Safety \\
\textbf{Description:} When two places $p$ and $p'$ enable isomorphic behaviors up to permutation of $p$ and $p'$, i.e. any transition consuming from $p$ has an equivalent but that consumes from $p'$, the tokens in $p$ and $p'$ enable the same future behaviors. We can fuse the two places into $p$, by redirecting arcs that feed $p'$ to instead feed $p$.  \\
\textbf{Definition:}
We let $v \equiv_{p\mid p'} v'$ denote equality under permutation of elements at index $p$ and $p'$ of two vectors $v$ and $v'$.

If $\exists p,p' \in \Places\setminus \mathit{Supp}, \Postset{p} \cap \Postset{p'} = \emptyset, \forall t \in \Postset{p},
\Pre(p,t)=1 \\
\land \exists t' \in \Postset{p'}, \Pre(t) \equiv_{p\mid p'} \Pre(t') \land \Post(t) \equiv_{p\mid p'} \Post(t') $, \\
then $\forall t \in \Preset{p'}$, update $\Post$ so that $\Post'(p,t) = \Post(p,t) + \Post(p',t)$, update initial marking to $m_0'(p)=m_0(p)+m_0(p')$, and discard $p'$ and transitions in $\Postset{p'}$.
\\
\textbf{Correctness:}
$\implies:$ Any firing sequence of the original net using transitions consuming from $p'$ still have an image using the transitions feeding from $p$. These two traces are observationally equivalent since neither $p$ nor $p'$ are in the support, so no behavior was lost. This transformation does not preserve the bounds on $p$'s marking however.
$\impliedby:$
The constraints on having only arcs with value $1$ feeding from $p$ and not having common output transitions feeding from both $p$ and $p'$ ensure there is no confusion problem for the
merged tokens in the resulting net; a token in $p'$ of the original net allowed exactly the same future behaviors (up to the image permutation) as any token in $p$ of the resulting net.
Merging the tokens into $p$ thus did not add more behaviors. Discarding transitions cannot add states. \\
\textbf{Notes:}
This test can be costly, but sufficient conditions for non-symmetry allow to limit complexity and prune the search space, e.g. we group places by number of output transitions, use a sparse "equality under permutation test"\ldots
The effect can be implemented as simply moving the tokens in $p'$ to $p$ and redirecting arcs to $p$, other rules will then discard the now constant place $p'$ and its outputs.

\subsection{Agglomeration rules}

Agglomeration in $p$ consists in replacing $p$ and its surrounding transitions (feeders and consumers) to build instead a transition for every element in the Cartesian product $\Preset{p} \times \Postset{p}$ that represents the effect of the sequence of firing a transition in $\Preset{p}$ then immediately a transition in $\Postset{p}$.
This "acceleration" of tokens in $p$ reduces interleaving in the state space, but can preserve properties of interest if $p$ is chosen correctly. This type of reduction has been heavily studied~\cite{HPP06,Laarman18} as it forms a common ground between structural reductions, partial order reductions and techniques that stem from transaction reduction. We present a very general version here, where we are agglomerating a subset $H \subseteq \Preset{p}$ of feeders of place $p$ to a subset $F \subseteq \Postset{p}$ of consumers of $p$.

\begin{definition}Let $p \in \Places$ and $H \subseteq \Preset{p}$ and $F \subseteq \Postset{p}$.\\
\textbf{(H,F)-agglomeration of place $\mathbf{p}$} : \\
$\forall h \in H, \forall f \in F, $ define a new transition $t$ such that\\
$$
\left\{
\begin{array}{@{}l@{}}
$Let $ k = \frac{\Post(p,h)}{\Pre(p,f)}, k \in \nat, k \geq 1,  \\
\Pre(t)=\Pre(h) + k \cdot \Pre(f) \land \Post(t)=\Post(h) + k \cdot \Post(f) \\
\end{array}\right.
$$
If $H=\Preset{p}$ and $F=\Postset{p}$ (complete agglomeration) discard transitions in both $H$ and $F$. \\
Else if $H \subset \Preset{p}$ discard transitions in $H$ only. \\
Else if $F \subset \Postset{p}$ discard transitions in $F$ only.
\end{definition}

\smallskip
Note the introduction of the $k$ factor, reflecting how many times $f$ can be fed by one firing of $h$. This factor must be a natural number for the agglomeration to be well defined (if there is no such $k$ we cannot apply an agglomeration). As a post processing, it is recommended to apply identity reduction Rule~$1$ to the set of newly created transitions, this set is much smaller than the full set of transitions but often contains duplicates. After a complete agglomeration the place $p$ is disconnected from the net and will be discarded by rule $9$ (constant place).

In the literature~\cite{EHPP05,HPP06}, the sets $H$ and $F$ are always the full preset and postset of the place, but our definition is more flexible when only part of the neighborhood of $p$ are stuttering transitions. We will use the term \textit{partial agglomeration} when the $H \subset \Preset{p}$ or $F \subset \Postset{p}$ strictly. In the rules given below, one of the two sets is always equal to the full preset or postset of $p$.
The definition of partial agglomeration was not present in~\cite{SRRPN20}.\smallskip

\rrule{Pre Agglomeration} \\
\textbf{Applicability:} Deadlock, Safety \\
\textbf{Description:}
Basically, we assert that once a stuttering transition $h \in \Preset{p}$ becomes enabled, it will stay enabled until some tokens move into the place $p$ by actually firing $h$. Transition $h$ cannot feed any other places, so the only behaviors it enables are continuations $f$ that feed from $p$. So we can always "delay" the firing of $h$ until it becomes relevant to enable an $f$ transition. We fuse the effect of tokens exiting $p$ using $f$ with its predecessor action $h$ by agglomerating in $p$. \\
\textbf{Definition:}

\vspace{2mm}
\begin{tabular}{@{}ll@{}}
$\exists p \in \Places\setminus \mathit{Supp}$  & $p$ not in support \\
$m_0(p)=0$ & initially unmarked \\
$\Preset{p} \cap \Postset{p} = \emptyset$ & distinct feeders and consumers \\
$\exists H \subseteq \Preset{p}, H \subseteq \Stutter{}$ & some feeders are stuttering \\
\end{tabular}

\begin{tabular}{ll}
$\forall h \in H,$ &  \\
& $\left\{
\begin{array}{@{}ll@{}}
\Post(p,h)=1, & $feed arc weights are one $ \\
\Postset{h}=\{p\} & p$ is the single output of $h \\
\exists p_1 \in \Places{}, \Post(p_1,h) < \Pre(p_1,h) & h$ is divergent free$ \\
\forall p_2 \in \Preset{h}, \Postset{p_2}=\{h\} & h$ is strongly quasi-persistent$ \\
\end{array}\right. $ \\
$\forall f \in \Postset{p},$  \\
&
$\left\{
\begin{array}{@{}ll@{}}
\Pre(p,f)=1 & $consume arc weights are one $ \\
\end{array}\right. $ \\
\end{tabular}

\medskip
Then perform an $(H,\Postset{p})$-agglomeration in $p$.\\
\textbf{Correctness:}  $\implies:$ Any sequence using an $h$ and an $f$ still has an image using the agglomerated transition $h.f$ in the position $f$ was found in the original sequence. Indeed, thanks to the quasi-persistent constraint, we know the necessary tokens tokens to fire $f$ will still be available after a delay when we want to fire $f$. Because $h$ is invisible and only feeds $p$ that is itself not in the support, delaying it does not lose any observable behaviors.  The divergent free constraint further ensures there are no runs that finish on $h^\omega$ (and thus we would never see $f$).
$\impliedby:$ Any state that is reachable in the new net by firing an agglomerate transition $h.f$ was already reachable by firing the sequence $h$ then $f$ in the original net, so no behavior is added. \\
\textbf{Notes:}
The terminology "strongly quasi-persistent" and "divergent free" is taken from~\cite{HPP06}. Pre agglomeration is one of the best known rules, this version is generalized to more than one feeder or consumer, and supports partial agglomeration. Of course it is desirable to have the largest possible $H$ set that satisfies the stuttering criterion, and when $H=\Preset{p}$ all transitions around $p$ are discarded, leading to also discard $p$ itself. When the agglomeration is only partial, we retain $p$ on the paths where it can be fed by an observed transition, but "accelerate" tokens just passing through $p$ on a stuttering execution.
\smallskip

\rrule{Post Agglomeration} \\
\textbf{Applicability:} Deadlock, Safety \\
\textbf{Description:}
Basically, we assert that once $p$ is marked, it fully controls its outputs, so the tokens arriving in $p$ necessarily have the choice of when and where they wish to go to. Provided $p$ is not in the support and at least some of its output transitions are invisible, we can fuse the effects of feeding $p$ with an immediate (unobservable) choice of what happens to those tokens after that.
 We fuse the effect of tokens entering $p$ using $h$ with an unobserved successor action $f$, and agglomerate around place $p$. \\
\textbf{Definition:}
If\vspace{2mm}

\begin{tabular}{@{}ll@{}}
$\exists p \in \Places\setminus \mathit{Supp}$  & $p$ not in support \\
$m_0(p)=0$ & initially unmarked \\
$\Preset{p} \cap \Postset{p} = \emptyset$ & distinct feeders and consumers \\
$\exists F \subseteq \Postset{p}, F \subseteq \Stutter{} $ & some consumers are stuttering \\
\end{tabular}

\begin{tabular}{ll}
$\forall f \in F,$  \\
$\left\{
\begin{array}{@{}ll@{}}
\Preset{f} = \{p\} & $no other inputs to $f$ $ \\
\forall h \in \Preset{p}, \Pre(p,h) / \Post(p,f) \in \nat & $natural ratio constraint$\\
\end{array}\right. $ \\
\end{tabular}\medskip

Then perform an $(\Preset{p},F)$-agglomeration in $p$.\\
\textbf{Correctness:}  $\implies:$ Any sequence using an $h$ still has an image using the agglomerated transition $h.f$ in the position $h$ was found in the original sequence, that leads to a state satisfying the same propositions since $f$ transitions stutter. Any sequence using an $f$ transition must also have an $h$ transition preceding it, and the same trace where the $f$ immediately follows the $h$ is feasible in both nets and leads to a state satisfying the same propositions. It is necessary that the $f$ transitions stutter so that moving them in the trace to immediately follow $h$ does not lead to observably different states.
$\impliedby:$ Any state that is reachable in the new net by firing an agglomerate transition $h.f$ was already reachable by firing the sequence $h$ then $f$ in the original net, so no behavior is added. \\
\textbf{Notes:}
Post agglomeration has been studied a lot in the literature e.g.~\cite{HPP06}, this version is generalized to an arbitrary number of consumers and feeders and a natural ratio constraint on arc weights, as well as partial agglomeration. As it is usually desirable to maximize the size of $F$, we simply take all stuttering transitions in $\Postset{p}$.
This procedure can grow the number of transitions when both $H$ and $F$ sets contain more than one transition, which becomes more likely as agglomeration rules are applied, and can lead to an explosion in the number of transitions of the net. In practice we refuse to agglomerate when the Cartesian product size is larger than $32$.

\rrule{Free Agglomeration} \\
\textbf{Applicability:} Safety \\
\textbf{Description:}
Basically, we assert that all transitions $h$ that feed $p$ \emph{only} feed $p$ and are invisible. It is possible that the original net lets $h$ fire but never enables a continuation $f$, these behaviors are lost since resulting $h.f$ is never enabled, making the rule only valid for safety. In the case of safety, firing $h$ makes the net lose tokens, allowing \textit{less} observable behaviors until a continuation $f \in \Postset{p}$ is fired, so the lost behavior leading to a dead end was not observable anyway. We agglomerate around $p$.
\eject

\hbox{}
\vspace*{-3mm}
\noindent \textbf{Definition:}

\vspace{2mm}
\begin{tabular}{@{}ll@{}}
$\exists p \in \Places\setminus \mathit{Supp}$  & $p$ not in support \\
$m_0(p)=0$ & initially unmarked \\
$\Preset{p} \cap \Postset{p} = \emptyset$ & distinct feeders and consumers \\
$\exists H \subseteq \Preset{p}, H \subseteq \Stutter{}$ & some feeders are stuttering \\
\begin{tabular}{ll}
$\forall h \in H,$ &  \\
& $\left\{
\begin{array}{@{}ll@{}}
\Postset{h}=\{p\} & p$ is the single output of $h \\
\Post(p,h)=1 & $feed arc weights is one $ \\
\end{array}\right. $ \\
$\forall f \in \Postset{p},$  \\
&
$\left\{
\begin{array}{@{}ll@{}}
\Pre(p,f)=1 & $consume arc weights is one $ \\
\end{array}\right. $ \\
\end{tabular}
\end{tabular}

\medskip
Then perform an $(H,\Postset{p})$-agglomeration in $p$.\\
\textbf{Correctness:}  $\implies:$
If there exists a sequence using one of the $f$ transitions in the original system, it must contain an $h$ that precedes the $f$. The trace would also be possible if we delay the $h$ to directly precede the $f$, because $h$ only stores tokens in $p$, it cannot causally serve to mark any other place than $p$, and since $h$ transitions are stuttering it leads to the same observable state in the new system. Traces that do not use an $f$ transition are not impacted. Sequences that use an $h$ but not an $f$ are no longer feasible, but because $h$ transitions stutter, the same sequence without the $h$ that is still possible in the new system would lead to the same observable states. So no observable behavior is lost as sequences and behaviors that are lost were not observable.
$\impliedby:$ Any state that is reachable in the new net by firing an agglomerate transition $h.f$ was already reachable by firing the sequence $h$ then $f$ in the original net, so no behavior is added. \\
\textbf{Notes:}
Free agglomeration is a new rule, original to~\cite{SRRPN20} that can be understood as relaxing conditions on pre-agglomeration in return for less property preservation. It is a reduction that may remove deadlocks, as it is no longer possible to fire $h$ without $f$, which forbids having tokens in $p$ that could potentially be stuck because no $f$ can fire. After firing $h$ the net is less powerful since we took tokens from it and placed them in $p$, these situations are no longer reachable. Due to its limited requirements, \textit{partial} free-agglomeration newly introduced in this paper is possible in many models as long as $p$ is unobserved.

\smallskip
\rrule{Controlling Marked Place} \\
\textbf{Applicability:} Deadlock, Safety \\
\textbf{Description:} A place $p$ that is initially marked and which is the only input of its single stuttering output transition $t$, can be emptied using $t$. Since $p$ controls its output, once it is emptied it will be post-agglomerable. \\
\textbf{Definition:}
If $\exists p \in \Places, \exists t \in \Stutter, \Postset{p} = \{t\}, \Preset{t} = \{p\}, p \notin \Postset{t}, \exists k \in \nat, m_0(p) = k \cdot \Pre(p,t)$, update $m_0' = m_0 + k \cdot \Effect(t)$. \\
\textbf{Correctness:} $\implies:$ Since $t$ is stuttering and only consumes from $p$, firing it at the beginning of any firing sequence will not change the truth value of the property in the reached state. $\impliedby:$ the new initial state was already reachable in the original model.  \\
\textbf{Notes:} This is the first time to our knowledge that a structural reduction rule involving token movement is proposed. This reduction
 may also consume some prefix behavior as long as a single choice is available.

\subsection{Graph-based reduction rules}

In this section we introduce a set of new rules that reason on a structural over approximation of the net behavior to quickly discard irrelevant behavior.
The main idea is to study variants of the token flow graph underlying the net to compute when sufficient conditions for a reduction are met.

In these graphs, we use places as nodes and add edges that partly abstract away the transitions. Different types of graphs are considered, all are abstractions of the structure of the net.  A graph is a tuple $G=(N,E)$ where nodes $N$ are places $N \subseteq \Places$ and edges $E$ in $\Places \times \Places$ are oriented.  We can notice these graphs are small, at most $|\Places{}|$ nodes, so these approaches are structural.

We consider that computing the prefix of a set of nodes, and computing strongly connected components (SCC) of the graph are both solved problems. The prefix of $S$
is the least fixed point of the equation $\forall s \in S, \exists s', (s',s) \in E \implies s' \in S$. The SCC of a graph form a partition of the nodes, where for any pair of nodes $(p,p')$ in a subset, there exists a path from $p$ to $p'$ and from $p'$ to $p$. The construction of the prefix is trivial; decomposition into SCC can be computed in linear time with Tarjan's algorithm.

\smallskip
\rrule{Free SCC} \\
\textbf{Applicability:} Deadlock, Safety \\
\textbf{Description:} Consider a set of places $P$ not in the support are linked
by elementary transitions (one input, one output). Tokens in any of these places can thus travel freely to any other place in this SCC. We can compute such SCC, and for each one replace all places in the SCC by a single "sum" place that represents it.\\
\textbf{Definition:}
We build a graph that contains a node for every place in $\Places{}\setminus \mathit{Supp}$ and an edge from $p$ to $p'$ iff.
$\exists t \in \Trans, \Preset{t}=\{p\} \land \Pre(p,t)=1, \Postset{t}=\{p'\} \land \Post(p',t)=1$.

For each SCC $S$ of size $2$ or more of this graph, we define a new place $p$ such that: $\forall t \in \Trans, \Pre(p,t) = \sum_{p' \in S} \Pre(p',t) \land \Post(p,t) = \sum_{p' \in S} \Post(p',t)$, and $m_0(p)=\sum_{p' \in S} m_0(p')$. Then we discard all places in the SCC $S$.\\
\textbf{Correctness:}  $\implies:$
Any scenario that required to mark one or more places  of the SCC is still feasible (more easily) using the "sum" place; no behavior has been removed.
$\impliedby:$ The sum place in fact represents any distribution of the tokens it contains within the places of the SCC in the original net. Because these markings were all reachable from one another in the original net, the use of the abstract "sum" place does not add any behavior. \\
\textbf{Notes:}
This powerful rule is computationally cheap, provides huge reductions, and is not covered by classical pre and post agglomerations. \cite{Srba19} has a similar rule limited to fusing two adjacent places linked by a pair of elementary transitions.
This rule (and a generalization of it) is presented using a  different formalization in~\cite{berthomieu19}.

\smallskip
\rrule{Prefix of Interest: Deadlock} \\
\textbf{Applicability:} Deadlock \\
\textbf{Description:} We consider a graph that represents all potential token flows, i.e. it has an edge from every input place of a transition to each of its output places. Only the presence of SCC (lakes) in this token flow graph can lead to an absence of deadlocks in the system ; if the net flows has no SCC (like a river), it must eventually must lose all its tokens and deadlock. In fact it must be the case that some tokens \emph{inevitably} end up in an SCC to prevent presence of deadlocks.
Tokens and places that are initially above (tributary or affluent) or in an SCC are relevant, as well as tokens that can help empty an SCC (they control floodgates). The rest of the net can simply be discarded.\\
\textbf{Definition:}
We build a graph $G$ that contains a node for every place in $\Places{}$ and an edge from $p$ to $p'$ iff. $\exists t \in \Trans, p \in \Preset{t} \land p' \in \Postset{t}$

We compute the set of non trivial SCC of this graph $G$: SCC of size two or more, or SCC consisting of a single place $p$ but only if it has a true self-loop $\exists t \in \Trans{}, \Preset{t}=\Postset{t}=\{p\} \land \Pre(p,t)=\Post(p,t)$.
We let $S$ contain the union of places in these non trivial SCC.
We add predecessors of output transitions of this set to the set,
$S \leftarrow S \cup \{ \Preset{t} \mid \exists p \in S, \exists t \in \Postset{p} \}$. This step is not iterated.
We then compute in the graph the nodes in the prefix of $S$, and add them to this set $S$.

We finally discard any places that do not belong to Prefix of Interest $S$, as well as any transition fed by such a place. Discard all places $p$, $p \notin S$, and transitions in $\Postset{p}$. \\
\textbf{Correctness:}  $\implies:$
The parts of the net that are removed inevitably led to a deadlock (ending in a place with no successor transitions or being consumed) for the tokens that entered them. These tokens now disappear immediately upon entering the suffix region, correctly capturing the fact this trace would eventually lead to a deadlock in the original net. So no deadlocks have been removed.
$\impliedby:$
Any scenario leading to a deadlock must now either empty the tokens in the SCC or consist in interlocking the tokens in the SCC. Such a scenario using only transitions that were preserved was already feasible in the original net, reaching a state from which a deadlock was inevitable once tokens had sufficiently progressed in the suffix of the net that was discarded.\\
\textbf{Notes:}
This very powerful rule is computationally cheap and provides huge reductions. The closest work we could find in the literature was related to program slicing rather than Petri nets. The main strength is that we ignore the structure of the discarded parts, letting us discard complex (not otherwise reducible) parts of the net. The case where $S$ is empty because the net contains no SCC is actually relatively common in the MCC and allows to quickly conclude.
\smallskip

\rrule{Prefix of Interest: Safety} \\
\textbf{Applicability:} Safety \\
\textbf{Description:} Consider the graph that represents all actual token flows, i.e. it has an edge from every input place $p$ of a transition $t$ to each of its output places $p'$ distinct from $p$, but only if $t$ is not just reading from $p'$.
This graph represents actual token movements and takes into account read arcs with an asymmetry. A transition consuming from $p_1$ to feed $p_2$ under the control of reading from $p_3$ would induce an edge from $p_1$ to $p_2$ and from $p_3$ to $p_2$, but not from $p_1$ to $p_3$. Indeed $p_1$ is not causally responsible for the marking in $p_3$ so it should not be in its prefix.

We start from places in the support of the property, which are interesting, as well as all predecessors of transitions consuming from them (these transitions are visible by definition). These places and their prefix in the graph are interesting, the rest of the net can simply be discarded.\\
\textbf{Definition:}
We build a graph $G$ that contains a node for every place in $\Places{}$ and an edge from $p$ to $p'$ iff.
$$\exists t \in \Trans, p \in \Preset{t} \land p' \in \Postset{t} \land p \neq p' \land \Pre(p',t) \neq \Post(p',t)$$

We let $S$ contain the support of the property $S=\mathit{Supp}$.

We add predecessors of output transitions of this set to the set, \\
$S \leftarrow S \cup \{ \Preset{t} \mid \exists p \in S, \exists t \in \Postset{p} \}$. This step is not iterated.

We then add any place in the prefix of $S$ to the interesting places $S$. We finally discard all places $p \in \Places{} \setminus S$, and for each of them the transitions in $\Postset{p}$. \\
\textbf{Correctness:}  $\implies:$
The parts of the net that are removed are necessarily stuttering effects, leading to more stuttering effects. The behavior that is discarded cannot causally influence
whether a given marking of the original net projected over the support is reachable or not. Any trace of the original system projected over the transitions that remain in
new net is still feasible and leads to a state having the same properties as the original net. So no observable behavior has been removed.
$\impliedby:$
Any trace of the new system is also feasible in the original net, and leads to a state satisfying the same properties as in the original net. So no behavior has been added.\\
\textbf{Notes:}
This very powerful rule is computationally cheap, provides huge reductions, and is not otherwise covered in the literature. Similarly to the rule for Deadlock, it can discard complex (not otherwise reducible) parts of the net. The refinement in the graph for read arcs allows to reduce parts of the net (including SCC) that are \emph{controlled by} the places of interest, but do not themselves actually feed or consume tokens from them.

\subsection{SMT-backed behavioral reduction rules}
\label{SMTRed}

Leveraging the over-approximation of the state space defined in Section~\ref{secSMT}, we now define reduction rules that test behavioral application conditions using this approximation.

\rrule{Implicit place} \\
\textbf{Applicability:} Deadlock, Safety \\
\textbf{Description:} An implicit place $p$ never restricts any transition $t$ in the net from firing: if $t$ is disabled it is because some other place is insufficiently marked, never because of $p$. Such a place is therefore not useful, and can be discarded from the net. \\
\textbf{Definition:}
Implicit place: a place $p$ is \emph{implicit} iff. for any transition $t$ that consumes from $p$, if $t$ is otherwise enabled, then $p$ is sufficiently marked to let $t$ fire. Formally,
$$
\forall m \in \Reach{}, \forall t \in \Postset{p}, (\forall p' \in \Preset{t}\setminus\{p\}, m(p') \geq \Pre(p',t)) \implies m(p) \geq \Pre(p,t)
$$

To use our SMT engine to determine if a place $p \in \Places{}\setminus \mathit{Supp}$ is assuredly implicit, we assert:
$$
\bigvee_{t \in \Postset{p}} ( m_p < \Pre(p,t) \land \bigwedge_{p' \in \Preset{t} \setminus\{p\}} m_{p'} \geq \Pre(p',t) )
$$

If the result in UNSAT, we have successfully proved $p$ is implicit and can discard it.

\textbf{Correctness:}  $\implies:$ Removing a place cannot remove any behavior.
$\impliedby:$ Removing the place $p$ does not add behavior since it could not actually disable any transition. \\
\textbf{Notes:}
The notion of implicit place and how to structurally or behaviorally characterize them is discussed at length in~\cite{colom99}, but appears already in~\cite{Berthelot85}.

We recommend to heuristically start by testing the places that have the most output transitions, as removing them has a larger impact on the net.
The order is important when two (or more) places are mutually implicit, so that each of them satisfies the criterion, but they share an output transition $t$ that only consumes from them (so both cannot be discarded).
\smallskip

\rrule{Structurally Dead Transition} \\
\textbf{Applicability:} Deadlock, Safety \\
\textbf{Description:} If in any reachable marking $t$ is disabled, it can never fire and we can discard $t$. \\
\textbf{Definition:}
For each transition $t \in \Trans$, we use our Safety procedure to try to prove the invariant "$t$ is disabled" negatively expressed by asserting:
$$
\bigwedge_{p \in \Preset{t}} m_p \geq \Pre(p,t)
$$

If the result is UNSAT, we have successfully proved $t$ is never enabled and can discard it. \\
\textbf{Correctness:}  $\implies:$
$t$ was never enabled even in the over-approximation we consider, therefore discarding it does not remove any behavior.
$\impliedby:$ Removing a transition cannot add any behavior.\\
\textbf{Notes:}
Because of the refined approximation of the state space we have, this test is quite strong in practice at removing otherwise reduction resistant parts of the net.

\section{Evaluation}

\subsection{Implementation}
The implementation of the algorithms described in this paper was done in Java and relies on Z3~\cite{z308} as SMT solver. The code is freely available under the terms of Gnu GPL, and distributed from \url{http://ddd.lip6.fr} as part of the ITS-tools.
Using sparse representations everywhere is critical; we work with transition based column sparse matrix (so preset and postset are sparse), and transpose them when working with places. The notations we used when defining the rules in this paper deliberately present immediate parallels with an efficient sparse implementation of markings and flow matrices. For instance, because we assume that $\forall p \in \Preset{t}$ is a sparse iteration we always prefer it to $\forall p \in \Places{}$ in rule definitions.

Our random explorer is also sparse, can restart (in particular if it reaches a deadlock, but not only), is more likely to fire a transition again if it is still enabled after one firing (encouraging to fully empty places), can be configured to prefer newly enabled transitions (pseudo DFS) or a contrario transitions that have been enabled a long time (pseudo BFS). It can be guided by a Parikh firing count vector, where only transitions with positive count in the vector are (pseudo randomly) chosen and counts decremented after each firing. For deadlock detection, it can also be configured to prefer successor states that have the least enabled events. These various heuristics are necessary as some states are exponentially unlikely to be reached by a pure random memory-less explorer. Variety in these heuristics where each one has a strong bias in one direction is thus desirable. After each restart we switch heuristic for the next run.

The setting of Section~\ref{secDefs} is rich enough to capture the problems given in the Model Checking Contest in the \textit{Deadlock},  \textit{ReachabilityFireability} and \textit{ReachabilityCardinality} examinations. We translate fireability of a transition $t$ to the state based predicate $m \geq \Pre(t)$. We also negate reachability properties where appropriate so that all properties are positive invariants that must hold on all states. To perform a reduction of a net and a set of safety properties, we iterate the following steps, simplifying properties and net as we progress:
\begin{enumerate}
    \item We perform a random run to see if we can visit any counter-example markings within a time bound.
    \item We perform structural reductions preserving the union of the support of remaining properties.
    \item We try to prove that the remaining properties hold using the SMT based procedure.
    \item If properties remain, we now have a candidate Parikh vector for each of them that we try to pseudo-randomly replay to contradict the properties that remain.
    \item If the computation has not progressed yet in this iteration, we apply the more costly SMT based structural reduction rules (see Sec.~\ref{SMTRed})
    \item If the problem has still not progressed in this iteration, we do a refined analysis to simplify atoms of the properties, reducing their support
    \item As long as at least one step in this process has made progress, and there remain properties to be checked, we iterate the procedure.
\end{enumerate}

Step~$6$ is trying to prove for every atomic proposition in every remaining property that the atom is invariant: its value in all states is the same as in the initial state. Any atom thus proved to be constant can then be replaced by its value in the initial state to simplify the properties and their support.
This procedure is also applicable to arbitrary temporal logic formulas, as shown in~\cite{Srba18}. We can thus in some cases even solve CTL and LTL logic formulas by reducing some of their atomic propositions to true or false.

\subsection{Experimental validation}

We analyze here the results of the Model Checking Contest 2020~\cite{mcc2019}\footnote{All data used in this section was collected and published by the competition organizers and made available at \url{https://mcc.lip6.fr}}, a competition for tools that can analyze Petri net specifications.
An examination consists in a model instance and either $16$ safety predicates (cardinality or fireability) or a single deadlock detection task. Models come from $103$ distinct families of Petri nets which are parameterized to build a total of $1229$ model instances\footnote{Due to a parse issue, our treatment of the colored variants of "Sudoku" model was incorrect. We thus filtered its $20$ instances from the analysis presented here to make $1209$ models.}. Of these $1209$ models, we able to parse and build a Petri net for $1195$\footnote{The missing $14$ are colored models that blow up to very large Petri net sizes ($10^7$ or more transitions), and one model that has more than $2^{32}$ tokens in a place. None of the other competitors was able to deal with these models either.}.

\textbf{The competition results:} We submitted two separate tools to the contest, ITS-tools and Lola (submitted with name "ITS-LoLa"). Both of them use the strategy presented in this paper, but ITS-Tools delegates residual models and properties to a decision diagram based engine\cite{tacas15}, while ITS-Lola delegates residual models to the model-checker Lola\cite{Wolf11} that uses a mixture of partial order reductions, ILP solving and explicit approaches. We used the versions of these tools submitted to the MCC in 2019, and which obtained respectively silver and bronze in deadlock detection and bronze and silver in the reachability category.

Thanks to the approach presented in this paper, in 2020 both of these submissions soared ahead of the competition scoring on deadlocks $17102$ and $17139$ when 2019 gold medalist Tapaal scored $14842$, and on reachability $35000$ and $34961$ when 2019 gold medalist Tapaal scored $33298$. The "best virtual score" $BVS$ which represents a virtual solver selected as the best one for any run is $35134$ on reachability\footnote{For deadlocks we are also very close to the BVS, but the value is not published on the website.} indicating that very few problems we didn't answer could be answered by another competitor. The very close scores of our two submissions is explained by the fact that the vast majority of answers was due to our new approach rather than to the back-end model-checker. In fact, removing the score contributed by the back-end Lola or ITS does not change the results, we would still get first position in both categories.

We can observe some variance between the ITS and the Lola runs in our approach due to the pseudo-random sampling but it does not exceed $10^{-3}$ with $1127$ vs $1125$ answers on deadlocks and $37243$ vs $37270$ answers on reachability. Interestingly the two back-end tools are able to solve a similar amount of the residual queries, $20$ vs $19$ answers on deadlocks, and $301$ vs $294$ answers on reachability were due to ITS-Tools and Lola respectively. So the back-end model-checkers contributed on deadlock $\approx 1.7 \%$ and on reachability only $\approx 0.8 \%$ of all answers computed\footnote{Because these results are so similar, the refined analysis below only uses the data collected from the ITS-Tools run.}.

\textbf{Deadlock detection:}
In this category our process has three ways of concluding for the 1127 answers computed~: 1. SMT over-approximation proves (UNSAT) that no deadlock state is feasible (in $246$ or $21.8 \%$ of answers), 2. The pseudo-random walk exhibits a deadlock (in $542$ or $48.1 \%$ of answers) 3. structural reductions may conclude using either Rule~19 (no SCC, deadlock is inevitable in $140$ or $12.4 \%$) or Rule~6 (source transition, no deadlock possible in $199$ or $17.7 \%$).

Drilling down into the $246$ SMT based answers, the process answered in the real domain to  $38 \%$ of queries, and in the natural domain to $62 \%$. Summing over reals and naturals, positive invariants were enough in $151$ queries, generalized invariants solved $54$, the state equation an additional $16$, and trap constraints added $25$ UNSAT results. Causal constraints did not obtain UNSAT on deadlock detection. The incremental way we add constraints and progress from reals to integers thus seems quite successful.

When looking at structural reductions the traces are harder to interpret, since they are not the first step in our process so we might not reach this step if we already found a counter-example. Still, $30 \%$ of the problems are directly solved at this step by rules 6 or 9. We measured which rules were applied at least once in a run, and report the percentage of runs the rule was used in. We find that $33 \%$ of runs used the future equivalent rule~13, $43 \%$ used pre-agglomeration rule~14, $67 \%$ used post-agglomeration rule~15, $28 \%$ could be reduced by the free SCC rule~18, $16 \%$ used the prefix rule~19, and SMT backed rules~21 and 22 were used respectively in $3 \%$ and $7 \%$ of runs. Of the $103$ families of nets, $86$ could be structurally reduced for deadlock detection.

\textbf{Reachability queries:}
In this category our process also had three ways of concluding for the 37243 answers computed~: 1. SMT over-approximation proves (UNSAT) that the invariant holds (in $6576$ or $17.6 \%$ of answers), 2. The pseudo-random walk exhibits a violation of the invariant (in $23428$ or $62.9 \%$ of answers) 3. structural reductions may conclude immediately (in $7239$ or $19.4 \%$) using rules~9 or 10 that identify places with a constant marking.

Drilling down into the $6576$ SMT based answers, the process answered in the real domain to  $84 \%$ of queries, and in the natural domain to $16 \%$. This is a strong argument in favor of trying reals first, instead of using a more complex ILP solver.
Summing over reals and naturals, positive invariants were enough in $3962$ queries, generalized invariants solved $1398$, the state equation an additional $1047$, trap constraints added $133$ results, and causal constraints contributed $36$ results.
It therefore appears that trying invariants before the more complex state equation is a very successful approach. The trap and causality constraints only marginally improve these results, but any result added is good as it avoids letting the random sampling look indefinitely for counter-examples that don't exist. The causal constraints are also very good at improving the feasability of the candidate Parikh vector when the response is SAT.

Again for structural reductions we measured the percentage of runs each rule was used in. Of the $103$ families of nets, $71$ could be structurally reduced by our rules which is much lower than for deadlock detection. In total $29.7 \%$ of the runs used one of the agglomeration variants of rules 14 to 16 and $17.6 \%$ used a partial version of these agglomeration. We also note that the new rule~16 (Free agglomeration) is used in $21.8 \%$ of runs which is quite significant.  The new graph based rules were also quite useful with $15.8 \%$ of runs benefiting from the prefix rule~20, and $11.8 \%$ benefiting from free SCC rule~18. Note that these percentages overlap, since a given run might benefit from several reduction rules.

For this examination we also looked into the decomposition of the $23428$ counter-examples found by the sampling component.
The pseudo-random and semi-heuristic runs found $92 \%$ of the counter-examples while the Parikh guided walk found $8 \%$. While this percentage may seem low, recall that the guided walk comes late in the process, after running SMT, so "easy" counter-examples are likely to have been found already.

\section{Conclusion}

The symbolic and structural approach presented in this paper combines over-approximation using an SMT solver, under-approximation by sampling with a random walk, and works with a system that is progressively simplified by property preserving structural reduction rules.

Structural approaches are a strong class of techniques to analyze Petri nets, that bypass state space explosion in many cases. Structural reductions are particularly appealing because any gain in structural complexity usually implies an exponential state space reduction. Symbolic manipulation of the state space constraints (invariants, state equation...) also has complexity related more to the size of the net than to its state-space.

The approach presented in this paper is complementary of other verification strategies as any problems incompletely solved can be submitted to another tool, the behavior for the given properties is preserved by the transformations.
Our approach is incomplete as it may be the case that an invariant holds, but the SMT constraints fail to prove it, and the sampling cannot prove it.
However in practice it solves a huge proportion of the queries, and despite being a new tool it was able to obtain gold medals in both Reachability and Deadlock categories of the model checking competition in 2020.

The choice of using an SMT based solver rather than a more classical ILP engine gives us flexibility and versatility, so that extending the refinement with new constraints is relatively easy. It enables an incremental approach starting from approximations in the real domain which despite being coarse are often enough by themselves.

We are currently investigating how to apply this approach to more complex problems such as model-checking stutter-insensitive CTL and LTL, as well as looking for new constraints or rules that we can add to our current set to reinforce its power.

\subsection*{Acknowledgements} The author would like to thank Emmanuel Paviot-Adet for many fruitful discussions on structural approaches in general, and on  systems of constraints that can be used to approximate the state space in particular.

\end{document}